\documentclass[prl,twocolumn,amsmath,amssymb]{revtex4}

\usepackage{graphicx}
\usepackage{bm}

\usepackage[varg]{txfonts}

\def\XXint#1#2#3{{\setbox0=\hbox{$#1{#2#3}{\int}$}
     \vcenter{\hbox{$#2#3$}}\kern-.5\wd0}}

\begin{document}

\title{Temperature dependent  spin susceptibility in a two-dimensional metal.} 

\author{V. M. Galitski}\affiliation{Kavli Institute for Theoretical Physics
and Physics Department, University of California, Santa Barbara, CA 93106}
\author{A. V. Chubukov}  \affiliation{Condensed Matter
  Theory Center, Department of Physics  University of Maryland,
  College Park, MD 20742-4111}
\author{S. Das Sarma} \affiliation{Condensed Matter
  Theory Center, Department of Physics  University of Maryland,
  College Park, MD 20742-4111}

\begin{abstract}
We consider a two-dimensional electron system with Coulomb interaction
between particles at a finite temperature $T$. We show that  the dynamic 
Kohn anomaly in the response function at $2k_F$ leads  to a non-analytic, 
linear-in-$T$ correction to the spin susceptibility, $\delta \chi (T) = AT$, 
same as in systems with short-range interaction.  
 We show that the singularity of the Coulomb  interaction at $q=0$
 does not invalidate the expansion of $A$ in powers of $r_s$, but makes
 the expansion non-analytic. We argue that the linear temperature
 dependence is consistent with the general structure of Landau theory and can be 
 viewed as originating  from the non-analytic component of the Landau function near the Fermi surface.    
\end{abstract}

\maketitle
\vspace*{-0.17in}

{\em Introduction ---} There has been substantial recent interest  in
the temperature dependence of various Fermi liquids properties 
 for both short-range and long-range interactions between particles
\cite{BKV,Chit,GSDS,num,CM,BKM}. 
The revival of interest in the problem is two-fold. 
On the experimental side, technical advances
 now allow one
 to measure the temperature dependence of the thermodynamic parameters
  such as specific heat and spin susceptibility 
 in ``classical'' 2D Fermi liquids with short-range interaction, 
 such as monolayers of $^3{\rm  He}$,
as well as study   two-dimensional semiconductor structures with long-range interaction and with 
relatively low Fermi temperatures ($\sim\! 1$\,K). On the theory side,
 the leading interaction corrections turn out to be 
 non-analytic  functions of temperature making the subject particularly important.

Na{\"\i}ve power counting arguments suggest that the temperature dependence
of any thermodynamic quantity, including the spin susceptibility and the
 specific heat coefficient $C(T)/T = \gamma$, should start with terms
quadratic in temperature.  
This conjecture  is based on the observation that a thermodynamic
quantity at a finite temperature typically can be written as
 $\int a(\varepsilon) n(\varepsilon)
d\varepsilon$, where $n(\varepsilon)$ is the Fermi distribution
function and $a(\varepsilon)$ is some function. If the latter is
smooth, the temperature dependence starts with a term of order $T^2$
\cite{Landau}. Such a temperature correction is called ``analytic.''
This is also consistent with the intuitive expectation of the one-to-one correspondence between the non-interacting Fermi gas and the interacting Fermi liquid since in the Fermi gas, the Sommerfeld expansion leads to simple quadratic temperature corrections.

However, the assumption about the analyticity of the functions
involved in the calculation of various thermodynamic properties of the
Fermi liquid is quite generally not justified because 
 in any Fermi liquid, the dynamic interaction between particles
 gives rise to  a non-analytic energy dependence of $a(\varepsilon)$.  
This leads to  temperature corrections which do not scale as
 $T^2$ and are therefore called  ``non-analytic.'' Collecting these non-analytic corrections is a subtle theoretical problem.

The subject of this paper is the 
 temperature correction to the spin susceptibility for a
 $2D$ system of fermions interacting via a long-range Coulomb interaction.
For Fermi systems with short-range interaction, perturbative 
 calculations for a model with a small $U(q)$ have demonstrated
 that $\delta \chi_s (T)$ is linear in $T$ and that the prefactor
 depends only on $U(2p_F)$. This result, however, was obtained under the
 assumption that for all $q$, the dimensionless Born parameter
 $u(q) =m U(q)/(2\pi)$ is  small. For Coulomb interaction this
 is obviously not the case, and one has to verify explicitly 
whether the linear 
dependence  $\delta \chi_s (T) \propto T$ still holds, and whether
 the prefactor can be expanded in $r_s$. The spin susceptibility 
 has recently been  measured at various $T$ in 
 Si inversion layers~\cite{R1,R2,R3,R4},
 and a quantitative theory is required to interpret
 the temperature dependence of the experimentally measured $\chi_s$. 
 In what follows, we compute the spin susceptibility 
in the perturbation theory and beyond, and relate the prefactor of the linear in $T$ term in an arbitrary Fermi liquid to the
 spin component of the quasiparticle scattering amplitude at the 
 scattering angle $\theta = \pi$. 

We also consider  in detail 
the relation  between the non-analytic $T$ dependence of the thermodynamic 
 parameters and  Landau Fermi liquid theory. The Landau theory operates
 with the  quasiparticle interaction function for the particles
 at the Fermi surface. In this theory, the spin susceptibility is independent
 of $T$, 
and is expressed via the particular partial component
 of the Landau function. The 
temperature corrections to the Fermi liquid theory  come from
 quasiparticles which are slightly off the Fermi surface. We demonstrate 
explicitly that non-analytic corrections to the spin susceptibility  can be viewed
 as originating from the non-analytic momentum dependence of the quasiparticle interaction function $f(p,p^\prime)$ 
at small deviations from the Fermi surface. We argue that
$f(p, p^\prime)$ is non-analytic in deviations from the Fermi surface, 
 and that this non-analyticity gives rise to  the emergence of the 
linear in $T$ terms in the $g-$factor and the spin susceptibility.

{\em Perturbation theory ---} 
Consider first the case when the Coulomb interaction 
 is weak at $q \sim p_F$, i.e., when $r_{\rm s} \ll 1$. 
The temperature correction to the spin susceptibility can be calculated
 either by explicitly evaluating the static particle-hole polarization bubble
 in a zero field, with insertions due to the interaction, or by evaluating the 
free energy in a finite magnetic field $H$ and then differentiating over $H$.
Either way, one obtains that,  to the leading (second) 
order in the interaction, the linear in $T$ term in the spin 
susceptibility comes from $2p_F$ processes, and 
\begin{equation}
\label{pt}
\delta \chi_s (T) = \chi_{\rm Pauli}\, \left( {r_{\rm s} \over 4} \right)^2 
{T \over E_{\rm F}},\,\,\,\,\mbox{ for }\, T/E_{\rm F} \ll 1,
\end{equation}
where $\chi_{\rm Pauli} = \mu_{\rm B}^2 m/\pi$ is the susceptibility of free fermions,
 and $r_s = \sqrt{2} m U(2p_F)/\pi$. 
The special role of $2p_F$ terms in the perturbation theory 
can be most easily understood  by
evaluating $\delta \chi_s (T)$ from the free energy. To second order in the interaction, the free energy $\Omega_2$ consists of two 
particle-hole bubbles connected by $U(q)$
\begin{equation}
\Omega_2 \propto U^2 (2p_F) T \sum_{\Omega_m} \sum_{\alpha\beta;\gamma \delta}
\int d^2 q \Pi^{\alpha,\beta} (q, \Omega_m) 
\Pi^{\gamma \delta} (q, \Omega_m),
\label{new_a}
\end{equation}
 where $\alpha,\beta, \gamma$ and $\delta$ are spin components of four
 fermions involved.
 Since the Coulomb interaction is spin independent, the spins of the two 
 fermions within each bubble are parallel (i.e., $\alpha =\beta$ and $\gamma = \delta$). 
However the spins in different bubbles can be either parallel or antiparallel  to each 
other. The singular $\delta \chi_s (T)$  comes from the 
antiparallel spin configuration between bubbles. 
For such spin orientation, fermions in each bubble 
 have  different  Fermi momenta $p^{+}_F$ and $p^{-}_F$ due to the 
Zeeman splitting. Near $2p_F$, this splitting is relevant as the   polarization bubble 
is non-analytic in both momentum and frequency. 
The non-analytic momentum dependence is normally associated with the Kohn anomaly and 
related Friedel oscillations. For the $T$ dependence of the spin
susceptibility, however,   one actually needs the dynamic polarization bubble. 
In 2D, the singular part of the polarization bubble
 behaves near $2p_F$ as  \cite{Stern}
\begin{eqnarray}
\label{Pi_sing}
 \Pi (q,\Omega_m)  \propto 
\sqrt{\left( {q \over 2 p_{\rm F}} + i {\Omega_m \over v_{\rm F}
    q} \right)^2 - 1} +
\sqrt{\left( {q \over 2 p_{\rm F}} - i {\Omega_m \over v_{\rm F}
    q} \right)^2 - 1}.
\end{eqnarray}
At small frequencies and $q < 2p_F$, this reduces to 
\begin{equation}
\Pi (q, \Omega_m) \propto \frac{|\Omega_m|}{\sqrt{2p_F -q}}.
\label{1}
\end{equation}
Integrating the product $\Pi^{++} (q, \Omega_m) \Pi^{--} (q, \Omega_m)$
 in (\ref{new_a}) over $q$, we find that the frequency dependence is not 
 analytic:
$\Omega_2 \propto U^2 (2p_F) T \sum_{\Omega_m} \Omega^2_m \log{[
\Omega^2_m + (\mu_B H)^2]}$. 
Evaluating the sum one finds that $\Omega_2$ contains a cross term $T H^2$~\cite{bec}. 
Differentiating over frequency, one then obtains $\delta \chi_s (T) \propto T$, as in (\ref{pt}). 
 
At the same time, the  potentially dangerous  small $q$ region, where 
 the Coulomb interaction is large, does not contribute 
to (\ref{pt}) 
(in this respect, our results differ from those in Ref 
~\cite{BKM}). The reason is that  at small $q$  
each of the  two  polarization bubbles contains only a non-singular, 
multiplicative dependence on the magnetic field; in two dimensions, this dependence
 comes through $v_F (H) = p_F (H)/m$ in $\Pi (q, \Omega_m) = (m/2\pi) \left[1 -
 |\Omega_m|/\sqrt{\Omega^2_M + (v_F q)^2}\right]$. This multiplicative 
dependence implies that the magnetic field only accounts for regular corrections in 
the form $(\mu_B H/E_F)^2$ for the $q=0$ piece in the free energy, i.e., no 
 crossed $T H^2$ term appears.

{\em Higher-order terms} --- As we just found, the second-order 
result for $\delta \chi_s (T)$ does not distinguish between short-range and 
 long-range interaction, i.e., the specifics of the Coulomb case does not show up. 
There is no guarantee, however, that this will remain so beyond the second order. 
Of particular interest is whether the divergence of the Coulomb interaction at 
$q=0$ affects the expansion of $\delta \chi_s (T) \propto T$ in powers of $r_s$. 
For this, we  computed the corrections to (\ref{pt}) from
 the third-order diagrams. We found that the divergent $U(0)$ still does not 
 directly contribute to the spin susceptibility, however the 
prefactor of the linear in $T$
 term gets modified due to vertex corrections to $U(2p_F)$. The 
 corrections involve $U(2p_F)$ itself and the momentum integrals of the 
 interaction potential. The most singular of these corrections
 accounts for the renormalization between $U(2p_F)$ and the spin component 
of the Landau function $\Gamma_s (\pi)$. This renormalization
 involves $T \sum_\omega \int d^2 p U(k-p) G(p,\omega) G(p-2k, \omega)$
 and yields the multiplicative correction to (\ref{pt}) in the form 
\begin{equation}
B = 1 + \frac{4}{\pi} \int_0^\pi d\theta ~u_\theta~ \cos{\frac{\theta}{2}}
 ~\log{\frac{\sqrt{1 + \sin{\theta/2}} +\sqrt{1 - \sin{\theta/2}}}{\sqrt{1 + \sin{\theta/2}}- \sqrt{1 - \sin{\theta/2}}}},
\label{2}
\end{equation}
where $u(\theta) = (m/2\pi) U(q = 2p_F \sin{\theta/2})$. One can easily verify 
that for short-range $u(\theta)$, the integral over $\theta$ converges, but 
 for the Coulomb interaction, when $u(\theta) \propto 1/\sin (\theta/2)$, 
the integral is confined to small $\theta$ and diverges as $\log^2$. 
The true divergence is indeed cut off by the screening effects.   
Still, it implies that
 in contrast to a short-range potential, 
 the prefactor for the linear in $T$ term in the spin susceptibility 
 in the 2D Coulomb system is
 non-analytic in $r_s$. 
Including screening in the usual way, we obtain  from (\ref{2}) 
  within logarithmic accuracy 
\begin{equation}
B = 1 + \frac{\sqrt{2} r_s}{\pi} \left[\log^2 r_s + O(\log r_s) +...\right],  
\label{3}
\end{equation}

In a generic Fermi liquid, the full linear in $T$ correction to the 
 spin susceptibility is evaluated in the same way as the correction to the specific 
heat~\cite{CM}. At each order of perturbation, one 
 selects two bubbles in which one keeps the  
 singular frequency dependence of $\Pi (q,\Omega_m)$, and 
 evaluate all other bubbles at $\Omega_0$. The two selected 
bubbles yield the $TH^2$  term, other bubbles contribute to the 
prefactor via the 
 renormalization of the $2k_F$ vertex.
  Extending the
 analysis in ~\cite{CM} to the spin susceptibility we obtain
 that the prefactor is expressed in terms of the spin component 
 of the full quasiparticle 
scattering amplitude at the scattering angle $\theta = \pi$.
In the explicit form,
\begin{equation}
\delta \chi_s (T) = \chi_{\rm Pauli} ~{T \over 2  E_{F}}
 \left[ \frac{m^*}{m}~F_{\rm sp} (\pi) \right]^2.
\label{4}
\end{equation}
Here $E_{F} = v_F p_F/2$ is the Fermi energy for free fermions, 
$m^*$ is the effective mass, and $F_{\rm sp} (\pi)$ is the spin component of the scattering amplitude. 
At weak coupling, $F_{\rm sp} (\pi) = - m U(2p_F)/(2\pi) = -r_s/(2\sqrt{2})$, and 
Eq. (\ref{4}) reduces to Eq. (\ref{pt}).  
In a generic Fermi liquid, $F_{\rm sp} (\pi) = \sum_{n=0} (-1)^n (2n +1)~ f_{{\rm sp},n}/(1 + f_{{\rm sp},n})$, 
where $f_{{\rm sp},n}$ are the partial spin components of the Landau function.

If the system is close to a ferromagnetic (Stoner) instability, $f_{{\rm sp},0} \approx -1$, and 
$F^2_{\rm sp} (\pi)$ can be well approximated by $f^2_{{\rm sp},0}/(1 + f_{{\rm sp},0})^2$ 
Then $\delta \chi_s (T) \approx (T/2E_{F}) \chi_s^2 / \chi_{\rm Pauli}$, where 
$\chi_s =  \chi_{\rm Pauli} (m^*/m) /(1 + f_{{\rm sp},0})$ is the spin susceptibility in a 
Fermi liquid at $T=0$. We did not analyze  higher-order terms in temperature, 
but based on the form of $\delta \chi_s (T)$, it is tempting to assume that    
 \begin{equation}
\chi^{-1}_s (T) = \chi^{-1}_s (T=0) - {T \over 2 E_{F}}  \chi^{-1}_{\rm Pauli}.
\label{5}
\end{equation}
If, on the contrary, the spin component of the scattering amplitude is
 small, but $m^*/m$ is arbitrary, as some studies suggest~\cite{R2},
 the same consideration yields
 \begin{equation}
\chi^{-1}_s (T) = \chi^{-1}_s (T=0) - {T \over 2 E_{F}} F^2_s (\pi) 
 \chi^{-1}_{\rm Pauli}.
\label{6}
\end{equation}
In  silicon inversion layer, $E_F \sim 6K$ for typical 
 densities~\cite{R2}.
 The susceptibility measurements 
have been reported for $T \sim 2-4K$. $\chi^{-1}_s (T)$ measured in units of 
 $\chi_{\rm Pauli}$ changes by about 20\% between $2K$ and $4K$~\cite{R2}. 
This would be
 consistent with Eq.~(\ref{5}), however the sign of the measured temperature correction 
is  opposite to that in (\ref{5}). 
Recent Shubnikov-deHaas  measurements, however, 
reported a much weaker, almost undetectable $T$ dependence of 
$\chi_s (T)$, from which  $\delta \chi_s (T)$ could not be 
extracted~~\cite{gershenson}. 
This much weaker effect would be more consistent with Eq.~(\ref{6}) if we
 assume that $F_{\rm sp} (\pi)$ remains small. More precise measurements of $\chi_s (T)$ 
are clearly called for to test our theoretical predictions.

{\em Extended Landau formalism} ---
We now consider in more detail  the physics behind the linear in $T$ 
dependence of the spin susceptibility. We argue that this term is actually consistent with 
the structure of Landau theory can be viewed as originating 
from the non-analytic structure of the Landau function near the Fermi surface. 

We remind that  the Landau function is the second variational 
derivative of the energy of the system with respect to the distribution function of
quasiparticles $n_\sigma({\bf p})$
\begin{equation}
\label{f}
f_{\sigma \sigma'} \left( {\bf p}, {\bf p}' \right)
= {\delta ^2 E \over \delta n_{\sigma}({\bf p}) \delta n_{\sigma}({\bf p}')}.
\end{equation}
The energy gain 
of a quasiparticle in a weak external magnetic field ${\bf H}$ is~\cite{Landau}
\begin{equation}
\label{de}
\delta \varepsilon ({\bf p}) = - \mu_{\rm B} 
\left( \bm{\sigma} {\bf H} \right) + {\rm Tr}_{\sigma'} \int 
f_{\sigma \sigma'} \left( {\bf p}, {\bf p}' \right)
{\partial n ({\bf p'}) \over \partial \varepsilon'}
 \delta \varepsilon ({\bf p'})
{d^2 {\bf p}' \over \left( 2 \pi \right)^2},
\end{equation}
where both $\delta \varepsilon$ and $\delta n$ are
matrices in the spin space.  The $g$-factor of a quasiparticle with
momentum ${\bf p}$ is defined by 
\begin{equation}
\label{g_def}
\delta \varepsilon ({\bf p}) = - {g  ({\bf p}) \mu_{\rm B} \over 2}
\left( {\bm{\sigma H}} \right),
\end{equation}
and the  spin susceptibility of a Fermi liquid is expressed as the momentum 
 integral over $ g({\bf p})$
\begin{equation}
\label{chi}
\chi = - {\mu_{\rm B}^2 \over 4} \int {\partial n ({\bf p}) \over
  \partial \varepsilon} g({\bf p}) {d^d {\bf p} \over \left( 2 \pi
  \right)^d},
\end{equation}

Eqs.~(\ref{de}) and (\ref{g_def}) determine the  
integral equation for the $g$-factor \cite{Landau}:
\begin{equation}
\label{g_int}
g({\bf p}) = 2  + {1 \over 2} \int 
f_{\rm sp} \left( {\bf p}, {\bf p}' \right)
{\partial n ({\bf p'}) \over \partial \varepsilon'}
 g ({\bf p'})
{d^d {\bf p}' \over \left( 2 \pi \right)^d}.
\end{equation}
Here $f_{\rm sp} $ is the spin component of the Landau function:
$\nu {\hat f} \left( {\bf p}, {\bf p}' \right) = 
f_c \hat I+ \bm{\sigma} \bm{\sigma}' f_{\rm sp} \left( {\bf p}, {\bf p}' \right)$,
where $\nu = m^*/\pi$.

 At zero temperature, the integration in (\ref{chi}) is confined to the Fermi surface.
At finite $T$,  the quasiparticles are allowed to deviate 
from the Fermi surface. Introducing $\xi = v_{\rm F} \left( p -
p_{\rm F} \right)$ and $\xi' = v_{\rm F} \left( p' - p_{\rm F}
\right)$, and assuming that $g({\bf p})$ depends on $\xi$ but not on the direction of ${\bf p}$, 
we re-write Eqs.~(\ref{g_int}) and (\ref{chi}) as
\begin{equation}
\label{g(xi)}
g(\xi) = 2  + \int 
\overline{f_{\rm sp}} \left( \xi,\xi' \right)
{\partial n  \over \partial \xi'}
g(\xi') d\xi'
\end{equation}
\vspace*{-0.1in}
and 
\vspace*{-0.1in}
\begin{equation}
\label{chi1}
\chi = -{\mu_{\rm B}^2 \nu \over 2} \int {\partial n \over
  \partial \xi} g(\xi) d \xi.
\end{equation}
Here  $\overline{f_{\rm sp}} \left( \xi, \xi' \right)$ 
is the interaction function averaged over the angle $\phi$ between the
momenta ${\bf p}$ and ${\bf p}'$:
$\overline{f_{\rm sp}} \left( \xi, \xi' \right) = \int_0^{2 \pi}
f_{\rm sp} \left( \xi, \xi'; \phi \right) {d \phi /(2 \pi)}$.

At $T=0$, only particles at the Fermi surface matter, and 
Eq. (\ref{g(xi)}) yields the well-known result 
$g(T=0) =  2/(1 + f_{{\rm sp},0})$
For calculations at a finite $T$, we
 need the solution at small but  finite $\xi$. 
To illustrate the appearance of singular terms, 
let us first study the structure of the finite temperature $f$-function
in the limit of weak interactions. The corresponding RPA correlation
energy can be written as follows \cite{Mahan}
\begin{equation}
\label{Ecor}
E_{\rm cor} = {\rm Re\,} \int {d e^2 \over e^2} \int {d\omega \over 2
  \pi i} \int {d^2 {\bf q} \over \left( 2 \pi \right)^2} 
\Pi(\omega, {\bf q}) \left[ V(\omega,{\bf q})
   - v({\bf q}) \right].
\end{equation}
 In Eq.~(\ref{Ecor}), $v(q)$ is the
bare Coulomb interaction in two dimensions and $V(\omega,{\bf q})$ is
the dynamically screened interaction $V(\omega,{\bf q}) = v({\bf q})
\left[ 1 - \Pi(\omega,{\bf q}) v({\bf q}) \right],^{-1}$.
The polarizability $\Pi(\omega,{\bf q})$ is defined as
\begin{equation}
\label{Pi}
\Pi(\omega,{\bf q}) = {\rm Tr_{\sigma} \,} \int {d \varepsilon \over 2
  \pi i} \int {d^2 {\bf p} \over \left( 2 \pi \right)^2} 
G_\sigma (\varepsilon,{\bf p} ) G_\sigma (\varepsilon + \omega,{\bf p} + {\bf q} )
\end{equation}
where $G_{\sigma} (\varepsilon,{\bf p})$ is the time-ordered Green's
function, which we write as a functional of the quasiparticle
distribution function:
\begin{equation}
\label{G}
G_{\sigma} (\varepsilon,{\bf p} ) = {n_\sigma ({\bf p}) \over \varepsilon
  - E_\sigma({\bf p}) - i0} + {1 - n_\sigma ({\bf p}) \over \varepsilon
  - E_\sigma ({\bf p}) + i0}. 
\end{equation}
Using the definition (\ref{f}) and Eqs.~(\ref{Ecor}), (\ref{Pi}), and
(\ref{G}), one can obtain the quasiparticle interaction function by a
straightforward evaluation of the derivatives with respect to $n_\sigma (p)$ \cite{Rice}.
At finite temperatures, we have to allow quasiparticles to depart 
from the Fermi surface. Keeping this in mind, we find that the
spin-dependent part of the Landau function is essentially the
dynamically screened Coulomb interaction, determined by the values of
momentum and energy transfer of the interacting quasiparticles:
\begin{eqnarray}
\label{fRPA}
f^{\rm (RPA)}_{\rm sp} ({\bf p}, {\bf p}') = -  
{\rm Re\,} V[E({\bf p}) - E({\bf p}'), {\bf p} - {\bf p}']. 
\end{eqnarray}

In the limit of low temperatures $T/E_{\rm F} \ll r_{\rm s} \ll 1$, we can
present the spin-dependent part of the Landau function 
$f({\bf p}, {\bf p}^\prime) = f_{\rm sp} (\xi,\xi^\prime;\phi)$ as
$f_{\rm sp} \left( \xi, \xi'; \phi \right) = 
f_{\rm reg} \left( \xi, \xi'; \phi \right) + 
f_{\rm sing} \left( \xi, \xi'; \phi \right)$, 
where the ``regular'' part is due to the statically screened Coulomb
interaction, and the ``singular'' part comes the singular {\it dynamic} 
 screening near $2k_F$. A small $\xi, \xi^\prime$, the
 singular part is a small correction to $f_{\rm reg}$ and
 can be written as
\begin{equation}
\label{fsing1}
f_{\rm sing} \left( \xi, \xi'; \phi \right) = - {1 \over \nu}
 \left[ f_{\rm reg} \left( \xi, \xi'; \phi \right)
\right]^2  
\Pi_{\rm sing} (\xi,\xi';\phi),
\end{equation}
where $\Pi_{\rm sing}$ is the dynamic part of $\Pi$, given by Eq. (\ref{Pi_sing}) at $\omega = E(p) - E(p^\prime)$. 

For actual calculations of the susceptibility, we will need the
 interaction function averaged over the angle $\phi$.
Performing the calculations, we find that 
 the singular part is non-analytic in the deviations from the Fermi surface.
\begin{equation}
\label{fsing}
\overline{f_{\rm sing}} \left( \xi, \xi' \right) = f_{{\rm sp},0} -
{(f_{{\rm sp},0})^2 \over 2} \left({m^*\over m}\right) 
{ \left| \xi - \xi' \right| \over E_{F} }+\ldots\, ,
\end{equation}
where the dots stay for regular terms in $\xi^2$ and $(\xi^\prime)^2$.
We explicitly verified that the non-analyticity in (\ref{fsing}) originates
 from the dynamic $2k_F$ Kohn anomaly.
This agrees with our diagrammatic analysis above. 

Substituting the non-analytic part of the Landau function into 
 the Eq.~(\ref{g(xi)}) for the $g$-factor, we obtain the following integral
 equation:
\begin{eqnarray}
\label{Int_g(T)}
\nonumber
g(\xi) =&&\!\!\!\!\!\! 2 +  f_{{\rm sp},0} \int\limits_{-E_{\rm F}}^{+\infty} 
g(\xi') {\partial n \over \partial \xi'} d\xi' \\
&-& {f^2_{{\rm sp},0}  \over 8 E_{F,0}} \left({m^*\over m}\right) 
  \int\limits_{-E_{\rm F}}^{+\infty} 
\left| \xi - \xi' \right| g(\xi') 
 {\partial n \over \partial \xi'} d\xi',
\end{eqnarray}
 This integral equation can be solved by iterations, 
using the zero temperature result $g(T=0)  = g^* = 2/(1 + f_{{\rm sp},0})$ as 
the first approximation. Performing the calculations, we obtain 
\begin{eqnarray}
\label{g(T)}
g(\xi,T) = && \!\!\!\!\!\!\! g^* \left[ 1 +
 f^2_{{\rm sp},0} {m^* \over m}~ \left( g^* -2 \right) { T \over
   8E_{\rm F}} \right] \nonumber \\ 
&+&  \!\!\!\!  f^2_{{\rm sp},0} {m^* \over m}~  
 {T \over {8 E_{F}}}
\left[  {\xi \over T} + 2 \ln\left(1 + e^{-\xi/T} \right)\right],
\end{eqnarray}
Using Eqs.~(\ref{chi1}) we then obtain for the spin
susceptibility
\begin{equation}
\label{chi_result}
\chi = {\mu_{\rm B}^2 \nu g^* \over 2} \left[ 1 -
 g^* f^2_{{\rm sp},0}  {m^* \over m} ~{T
  \over {8 E_{\rm F}}} \right].
\end{equation}
This agrees with the result of the diagrammatic
 treatment, Eq.~(\ref{4}), for the case when the full scattering amplitude
$F_{\rm sp} (\pi)$ can be approximated by its zeroth partial 
component $F^2_{\rm sp} \approx f^2_{{\rm sp},0}/(1 + f_{{\rm  sp},0})^2$. This approximation is 
 implicit in the RPA formalism. In a more generic analysis, one indeed
 should recover the full scattering amplitude.

We see therefore that the linear in $T$
 correction to the spin susceptibility  can be understood as originating
 from the non-analytic momentum dependence of the Landau function 
 at small deviations from the Fermi surface. 
 This
 non-analytic momentum dependence $f_{\rm sing} \propto |\xi - \xi^\prime|$
 is the fundamental consequence of the dynamic $2p_F$ 
Kohn anomaly in a generic Fermi liquid.



To conclude, we considered the temperature dependence 
 of the spin susceptibility in a 2D electron system with Coulomb interaction.
We found that the leading temperature correction $\delta \chi_s (T)$ is linear 
in $T$ and comes from the $2p_F$ singularity in the dynamical response function
 (the dynamic Kohn anomaly). The origin of the effect is the same as in systems 
with short-range interaction. However, for Coulomb interaction, 
 the prefactor of the $O(T)$ term is itself non-analytic in $r_s$. 
We also analyzed the emergence of the $O(T)$ term in  $\delta \chi_s (T)$
 by extending the Landau formalism to finite $T$. We demonstrated that within
 this approach, the non-analytic temperature dependence of the susceptibility originates from the non-analytic momentum dependence of the Landau function at small deviations from the Fermi surface. The experimental verification of the linear in $T$ dependence of the spin susceptibility is clearly called for.

{\em Acknowledgments ---} We thank D. Maslov for useful conversations. 
VG was supported by the David and Lucile Packard foundation.
SDS was supported by the US-ONR, LPS, and DARPA.  AVC was supported 
 by NSF DMR 0240238.

\vspace*{-0.2in} \bibliography{chi}

\end{document}